\pdfoutput=1
\documentclass[journal]{IEEEtran}
\usepackage{graphicx}
\usepackage{dcolumn}
\usepackage{bm}
\newcommand{\forb}{\mbox{FO2-Rb}}
\newcommand{\focs}{\mbox{FO2-Cs}}

\begin{document}
%
\title{Contributing to TAI with a Secondary Representation of the SI Second}

\author{J.~Gu\'{e}na, M.~Abgrall, A.~Clairon and S.~Bize
\thanks{Manuscript received 8 November 2013, revised 9 December 2013

Accepted for publication 13 December, 2013

Published 29 January 2014}
\thanks{Authors are affiliated to:
LNE-SYRTE, Observatoire de Paris, CNRS, UPMC, 61 avenue de l'Observatoire, 75014 Paris, France.

Email: jocelyne.guena@obspm.fr}
}

\maketitle

\begin{abstract}
We report the first contribution to the international atomic time (TAI) based on a secondary representation of the SI second. This work is done with the LNE-SYRTE {\forb} fountain frequency standard using the $^{87}$Rb ground state hyperfine transition.
 We describe {\forb} and how it is connected to local and international time scales. We report on local measurements of this frequency standard in the SI system, \emph{i.e.} against primary frequency standards, down to a fractional uncertainty of $4.4\times 10^{-16}$, and on the establishment of the recommended value for the $^{87}$Rb hyperfine transition by the CIPM. We also report on the process that led to the participation of the {\forb} frequency standard to \emph{Circular~T} and to the elaboration of TAI.
 This participation enables us to demonstrate absolute frequency measurement directly in terms of the SI second realized by the TAI ensemble with a statistical uncertainty of $1.1\times 10^{-16}$, therefore at the limit allowed by the accuracy of primary frequency standards. This work constitutes a demonstration of how other secondary representations, based on optical transitions, could also be used for TAI, and an investigation of a number of issues relevant to a future redefinition of the SI second.
\end{abstract}


\section{Introduction}
The second of the international system of units (SI) is defined using the ground state hyperfine transition of the caesium 133 atom since 1967 \cite{CGPM1967}\cite{Terrien1968}. The accuracy of the widely used international atomic time (TAI) is now provided by atomic caesium fountains used as primary frequency standards (PFSs), which give the most accurate realization of the SI second. In 2001, considering the development of new frequency standards \emph{``which could eventually be considered as the basis for a new definition of the second''}, the Consultative Committee for Time and Frequency (CCTF) of the Comit\'e International des Poids et Mesures (CIPM) recommended that \emph{``a list of such secondary representations of the second be established''} \cite{CCTF2001}. In 2004, the CCTF proposed a first secondary representation of the second \cite{CCTF2004} which was then adopted by the CIPM \cite{CIPM2007}. By 2012, the list had grown to 8 secondary representations \cite{CCTF2012}\cite{CIPM2013}.

Secondary representations of the SI second (SRS) are transitions which are used to realize frequency standards with excellent uncertainties,
and which are measured in the SI system with accuracies close to the limit of Cs fountains 
\cite{Rosenband2007}-\cite{Guena2010}~\footnote{See the list of Secondary Representations of the SI second and of other recommended values of standard frequencies on the BIPM website. Available from: http://www.bipm.org/en/publications/mep.html.}. Several of these frequency standards based on an SRS have already achieved estimated uncertainties well beyond those of PFSs,
which opens the inviting prospect of a redefinition of the SI second \cite{Gill2011}\cite{Riehle2012}. Establishing and improving the list of SRSs is crucial but is only the first step toward a redefinition.
Pending questions must be answered before such a redefinition can occur. For most standards based on an SRS, there is still a considerable challenge to reach the level of reliability and operability needed for these standards to practically improve dissemination via international timekeeping beyond the current state-of-the-art. This challenge in fact concerns not only the standards themselves but also other key elements of the timekeeping process such as means of remote comparisons and local oscillators \cite{Parker2012}. Furthermore, it is crucial to have mechanisms that will ensure the best possible continuity in the dissemination of the SI second and in timekeeping across the change of definition,
despite the necessarily non perfectly identical primary and secondary realizations in different institutes. Also, there is so far a limited number of stringent comparisons between secondary frequency standards (SFSs) based on the same transition to test their uncertainty well below that of atomic fountains. It is highly desirable that more of these measurements be done, ideally between different institutes. Last but not least, international agreement has to be reached on choices that will be made.

In this article, we report the first contribution to TAI using a secondary representation of the SI second. We describe the frequency standard itself, which is based on the $^{87}$Rb ground state hyperfine transition used in the FO2 dual atomic fountain. We will refer to this standard as {\forb}. We discuss the uncertainty and stability obtained with {\forb} and describe the frequency metrology chain between {\forb} and TAI. Moreover, we report on our highly accurate measurements of the $^{87}$Rb hyperfine transition against PFSs and on the establishment of the recommended value for the $^{87}$Rb hyperfine transition based on these measurements. We describe the process of submitting {\forb} data to the BIPM, the inclusion of these data in \emph{Circular~T}, and the stringent comparisons of {\forb} to the TAI ensemble. This work contributes to investigating the above pending questions. It demonstrates how other secondary representations based on optical transitions could also contribute to TAI and how the recommended values could be
compared to the SI second as realized by the TAI ensemble. Arguably, several of these steps are necessary before a secondary representation can serve as a basis for redefining the SI second.

\section{The {\forb} atomic fountain}

\begin{figure}[h]
\includegraphics [width= \linewidth]{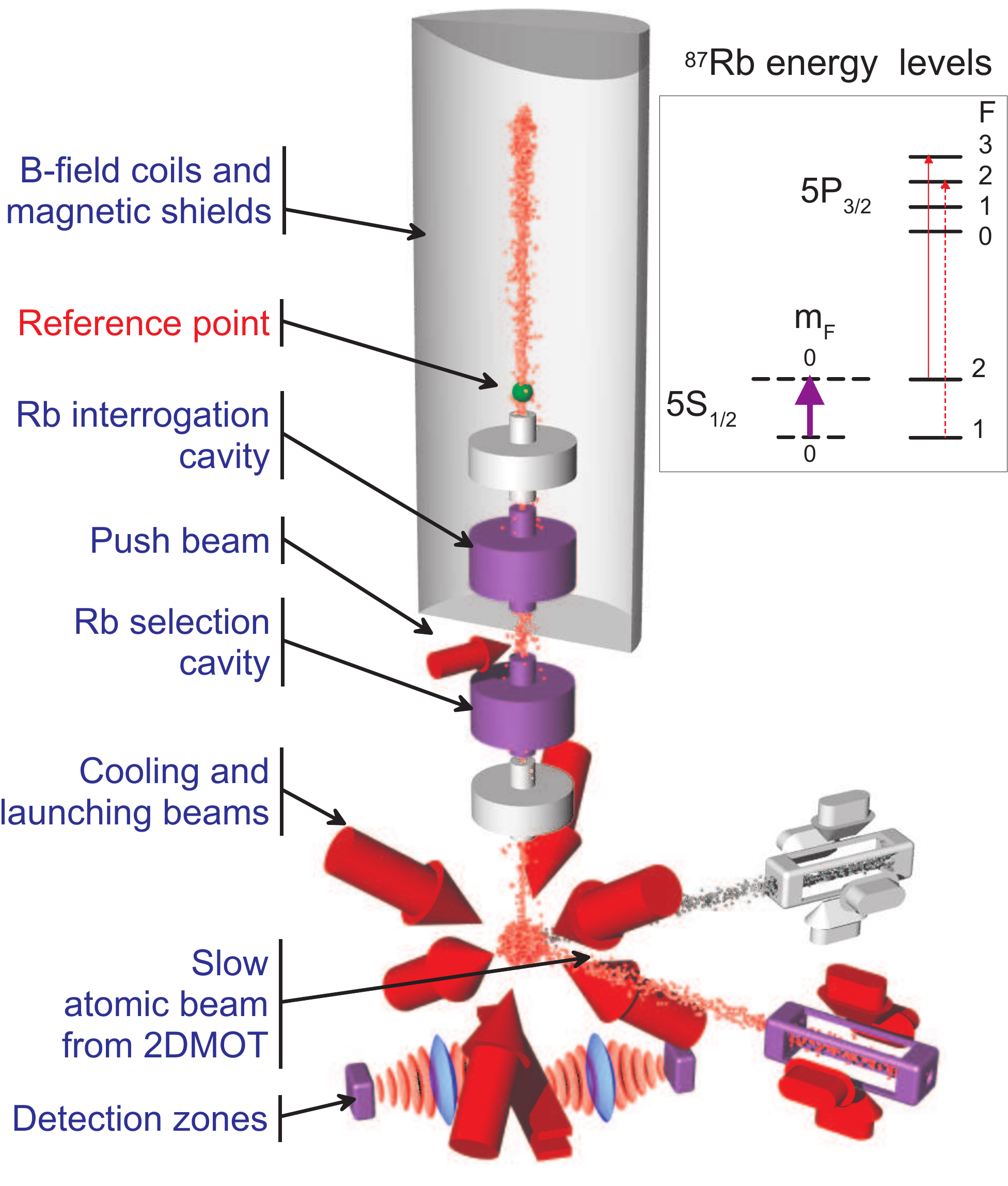}
\caption{Schematic view of the FO2 dual Rb$\&$Cs fountain. The 852~nm light and the 780~nm light for $^{133}$Cs and $^{87}$Rb atoms are superimposed in each of the 9 laser beams with dichroic beam splitters. $^{133}$Cs and $^{87}$Rb atoms are captured, launched, state selected, probed and detected simultaneously at each fountain cycle of $1.6$~s. Two independent 2 dimensional magneto-optic traps are used to load the optical molasses with the two species. The reference point of {\forb} is located 0.146~m above the Rb interrogation cavity (see Sect. \ref{subsec_redshift}). Not shown are two magnetic shields surrounding the whole vacuum system, including the molasses and detection region. In insert the purple thick arrow shows the $^{87}$Rb ground state hyperfine transition which is used in this work. The thin red arrows show optical transitions used for laser cooling and detecting Rb atoms.}
\label{fig_FontaineRbCs}
\end{figure}

The {\forb} fountain is the rubidium part of the dual fountain FO2 schematized in Fig. \ref{fig_FontaineRbCs}, which uses both Rb and Cs atoms simultaneously. In the following, we focus on the description of the rubidium part.
The source of $^{87}$Rb atoms is a slow beam emerging from a 2 dimensional magneto-optic trap (2DMOT) which eliminates both the unwanted $72.2\%$ abundant $^{85}$Rb isotope and the $^{87}$Rb background in the main vacuum chamber. Atoms are captured (typical loading time of 600~ms) and laser-cooled in a Lin $\perp$ Lin
optical molasses in the (1,1,1) configuration~\footnote{
An optical molasses in the (1,1,1) configuration is made of 3 pairs of counter-propagating laser beams aligned along the axes of a 3-dimensional orthonormal basis where the (1,1,1) direction is along the vertical direction. The Lin $\perp$ Lin configuration refers to the case where laser beams are linearly polarized, with orthogonal polarizations between the counter-propagating beams.} sketched in Fig. \ref{fig_FontaineRbCs}.
The light at 780~nm for all main beams (molasses, state selection and detection) is fiber-coupled to the vacuum system using specific collimators that ensure proper alignment at the few $100~\mu$rad level. A dichroic plate in these collimators combines the light at 780~nm with the light at 852~nm for the caesium part of the fountain \cite{Guena2010}. The molasses beam diameter (at 1/e$^{2}$) is $\simeq$~26~mm. Three home-built extended cavity diode lasers fitted with a narrow-band interference filter for wavelength selection \cite{Baillard2006} are employed to generate the 780~nm light. The detection laser is locked by saturated absorption spectroscopy to one hyperfine component of the D2 line using the modulation transfer technique providing a high quality lock without modulating the laser frequency \cite{Shirley1982}.
The two other seed lasers (repumper and cooling laser) are frequency-stabilized using their beat notes with the detection laser, used as a reference.
The necessary amount of optical power is obtained using tapered semi-conductor laser amplifiers for the 2DMOT and the molasses beams ($\sim$~80mW in total for the 2DMOT, $\sim$~10~mW for each of the 6 molasses beams), and by injection-locking another laser diode for detection.
Atoms gathered in the upper $F=2$ hyperfine level of the $^{2}S_{1/2}$ ground state are launched upwards using the moving molasses technique at a velocity of $4.158$~m.s$^{-1}$ (launch height 0.881~m above the center of the molasses) and a temperature of $1.6~\mu$K, corresponding to a rms velocity of $\sim 1.2$~cm.s$^{-1}$ or $\sim 2$ recoils along one direction. They are then selected in the $|F=1,m_F=0\rangle$ Zeeman sub-level of the lower $F=1$ hyperfine ground state by a 2~ms microwave pulse in the state-selection cavity (diameter 58~mm, length 62~mm) located 0.139~m above the molasses center. Atoms remaining in the $F=2$ state are pushed away by a light pulse. The interrogation microwave cavity located 0.442 m above the molasses is a TE$_{011}$ copper resonator (lower part of the Cs/Rb dual cavity, Fig.3 of \cite{Guena2010}) fitted with two independent microwave feedthroughs. The loaded quality factor of this cavity is $\sim 6000$. Atoms in the two clock states are detected $0.13$~m below the molasses by time-resolved laser-induced fluorescence.
In the simultaneous operation of {\forb} and {\focs}, the launch height for the Rb cloud is slightly less than for the Cs cloud so that the 2 clouds can be selectively detected with no time overlap \cite{Guena2010}.
The transition probability from the $|F=1,m_F=0\rangle$ clock state to $|F=2,m_F=0\rangle$ is alternately measured at the two sides of the central Ramsey fringe (full width at half maximum 0.82~Hz, contrast $>99.5\%$) to lock the interrogation microwave signal to the clock transition at 6.8~GHz. Further details about the FO2 fountain setup can be found in \cite{Guena2010}, \cite{Guena2012}-\cite{Chapelet2008}.
A key factor to achieve low fractional frequency instability is the noise level of the atomic detection. Our detection system has a cycle-to-cycle noise floor equivalent to $\sim 400$ atoms. The highest detected atom number is $2.4\times 10^{6}$, so that the measurement of the transition probability is predominantly limited by quantum projection noise \cite{Santarelli1999}. We have measured fractional frequency instability of $3\times 10^{-14}\tau^{-1/2}$ in agreement at the $5\%$ level with the projection noise limit calculated for the calibrated atom number.

\section{Uncertainties in the {\forb} frequency standard} \label{sec_accuracy}

Table \ref{tab_accuracy} gives the uncertainty budget of the {\forb} frequency standard as of 2012. Generally, physical effects, magnitudes of corrections and their Type~B uncertainties, are similar to those found in caesium. Below, we discuss each of these effects. Corrections and uncertainties are subjected to small updates, due to changes in the environment, in the atom number or in the measurement duration.

\begin{table}
\begin{center}
\caption{Uncertainty budget of the {\forb} secondary frequency standard}
\label{tab_accuracy}
\begin{tabular}{lcc}
\hline \hline 
Physical origin of the shift                                       & Correction   & Uncertainty  \\
\hline 
{\small Quadratic Zeeman}               & $-3465.5$&$0.7$  \\
{\small Blackbody radiation}                  &   $124.2$&$1.4$\\
{\small Collisions and cavity pulling}        &   $8.0$&$2.1$\\
{\small Distributed cavity phase}       & $0.4$& $1.0$   \\
{\small Microwave lensing}                     &  $-0.7$&$0.7$   \\
{\small Spectral purity \& leakage}            & 0  &$0.5$ \\
{\small Ramsey \& Rabi pulling}                & 0 &$0.1$  \\
{\small Relativistic effects}            & 0 & $0.05$ \\
{\small Background collisions}                 & 0  &$1.0$ \\
\hline 
{\small Total}                               &  $-3335.0$&$3.2$  \\
\hline \hline
\end{tabular}
\end{center}
{The table gives the fractional frequency correction and its Type~B uncertainty for each systematic shift, in units of $10^{-16}$. The total uncertainty is the quadratic sum of all uncertainties.}
\end{table}

\subsection{Second order Zeeman shift}\label{subsec_ZeemanShift}

\begin{figure}[h]
\includegraphics [angle=-0, width=\linewidth]{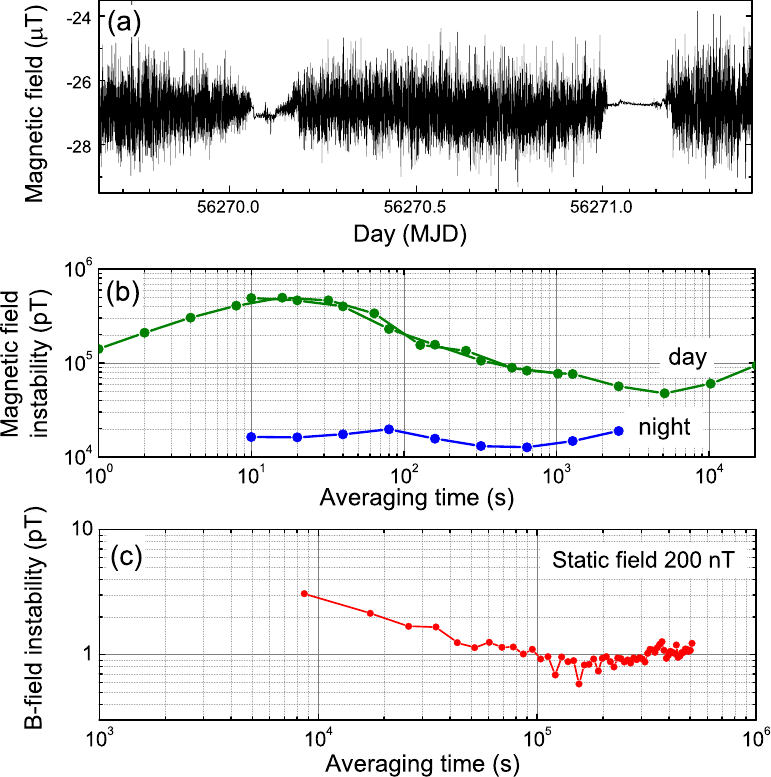}
\caption{(a) Vertical component of the external ambient magnetic field in FO2 laboratory measured with a fluxgate magnetometer over two days. (b) Instability of this magnetic field during days and nights. (c) Instability of the internal static magnetic field sampled every hour at the nominal launch height of 0.881~m during the operation of {\forb} as a frequency standard over several months.}
\label{fig_Bstability}
\end{figure}

\begin{figure}[h]
\includegraphics [width= \linewidth]{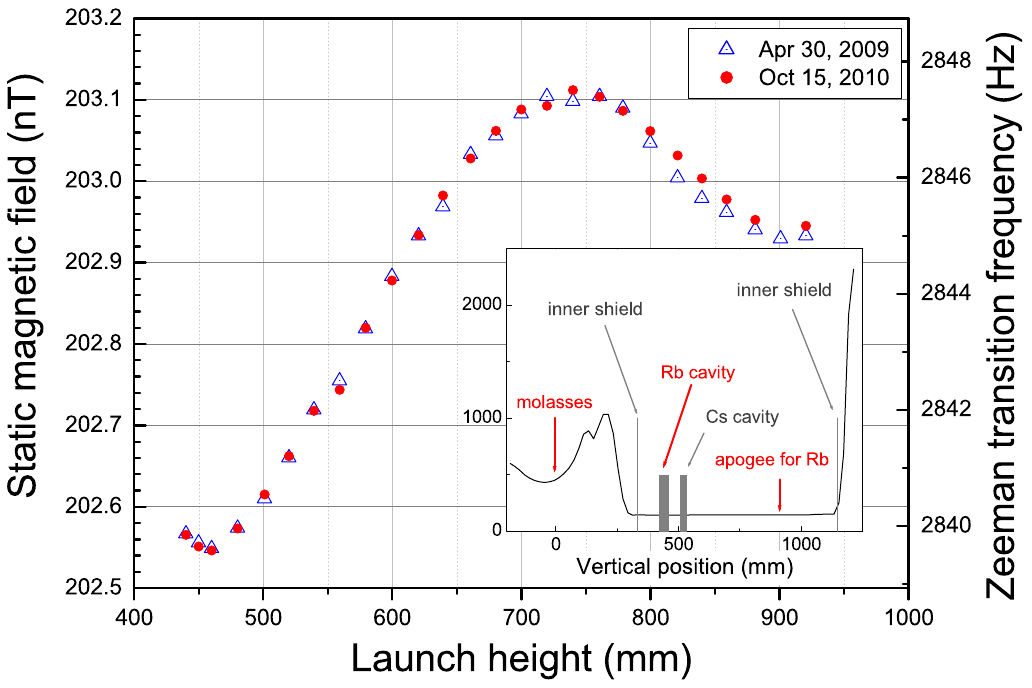}
\caption{Static magnetic field map in the interrogation region of {\forb} probed by spectroscopy of the Zeeman $|F=1,m_F=1\rangle \longrightarrow |F=2,m_F=1\rangle$ component of the $^{87}$Rb ground state hyperfine transition. The reference for the launch height is the center of the molasses. Inset: Profile of the vertical component of the field initially measured in FO2 with a fluxgate magnetometer.}
\label{fig_Bmap}
\end{figure}

In $^{87}$Rb the second-order Zeeman coefficient for the clock transition frequency is given by $K_{Z2}=5.75146 \times 10^{10}$~Hz.T$^{-2}$ corresponding to a fractional shift of $8.4151\times 10^{-12}\mu$T$^{-2}$ ($\sim1.81\times$ larger than for Cs). The first order Zeeman coefficient for the $|F=1,m_F=1\rangle \longrightarrow |F=2,m_F=1\rangle$ transition is given by $K_{Z1}=1.40195 \times 10^{10}$~Hz.T$^{-1}$ ($\sim2\times$ larger than for Cs). The coefficients are determined using the Breit-Rabi formulae (see, for example, \cite{Vanier1989}) with g-factors of \cite{Arimondo1977}.
In FO2, there are two magnetic shields (not shown in Fig.~\ref{fig_FontaineRbCs}) that surround the whole vacuum system, including the molasses and detection region. The vertical magnetic field component is measured with a fluxgate magnetometer in the molasses region and is actively stabilized with a set of horizontal coils distributed over the height of the fountain. Three additional shields with endcaps surround the interrogation region to further attenuate ambient magnetic field fluctuations which are quite large at Observatoire de Paris due to the proximity of the metropolitan transportation system. The upper plots of Fig. \ref{fig_Bstability} display these fluctuations exhibiting the change between day and night. The overall shielding factor is $\sim 2.5 \times 10^{5}$. Inside the innermost shield, a solenoid (length 815~mm, pitch 10~mm) with back and forth winding over the whole length, supplemented with a set of 4 compensation coils provides a static magnetic field of $\sim200$~nT.
The magnetic field is homogeneous to $10^{-3}$ as can be seen in Fig. \ref{fig_Bmap} which shows two field maps recorded with {\forb} at a two years interval using the spectroscopy of the $|F=1,m_F=1\rangle \longrightarrow |F=2,m_F=1\rangle$ field sensitive transition.
The inset in Fig. \ref{fig_Bmap}
shows the profile of the vertical component of the magnetic field measured early on in FO2 using a fluxgate sensor to verify the absence of zero crossing, of large gradients on the atomic trajectories and of field reversal at the entrance of the interrogation region. This is important to avoid Majorana (spin flip) transitions and therefore ensure control of the quantization axis in the fountain.
Fig. \ref{fig_Bstability}(c) shows the instability of the static magnetic field probed via Zeeman spectroscopy performed at the nominal launch height (0.881~m) every hour during operation of FO2 as a frequency standard. The field is stable to $\sim 1$~pT ($\sim 5\times10^{-6}$ fractionally) over months.

The uncertainty in Table \ref{tab_accuracy} accounts for temporal fluctuations and for the statistical uncertainty of  the magnetic field measurement. The impact of the inhomogeneities of the static magnetic field, less than 0.6 nT, is negligible.

\subsection{Blackbody radiation shift} \label{subsec_BBR}

\begin{figure}[h]
\includegraphics [width= \linewidth]{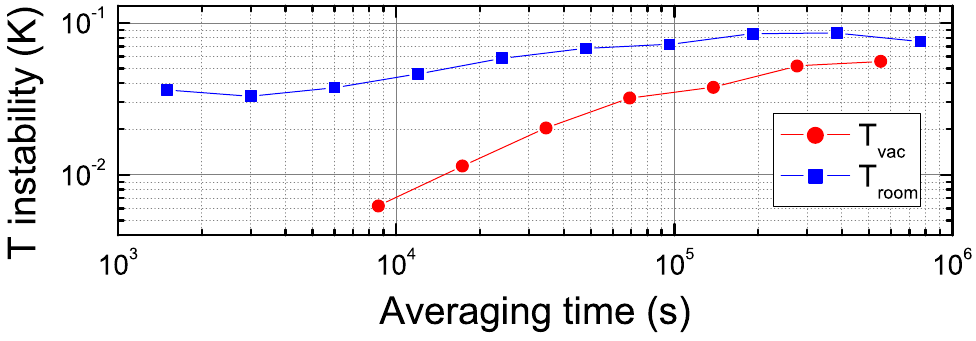}
\caption{Room (T$_{\mathrm{room}}$) and vacuum chamber (T$_{\mathrm{vac}}$) temperature instability over 2 months. T$_{\mathrm{vac}}$ is the temperature monitored by the central sensor (see Sect. \ref{subsec_BBR}).}
\label{fig_Tstability}
\end{figure}

The blackbody radiation (BBR) shift at an absolute  temperature T can be written as :
\begin{equation}\label{Eq_BBR}
  \beta \left(\frac{T}{T_{0}}\right)^{4}\times \left[1+ \epsilon \times\left(\frac{T}{T_{0}}\right)^{2}\right]
\end{equation}
with $T_{0}$=300~K. For Rb, two independent high accuracy \emph{ab initio} calculations give $\beta _{\mathrm{Rb}}^{\mathrm{theo}}=-1.2557\times 10^{-14}$ with 1 and 0.3 percent uncertainty respectively \cite{Angstmann2006}\cite{Safronova2010a}. The dynamic correction $\epsilon$ is also estimated in \cite{Angstmann2006}: $\epsilon_{\mathrm{Rb}}^{\mathrm{theo}}=0.011$. $\beta$ is related to the scalar Stark shift coefficient $k^{s}$ of the clock transition in a static electric field:
$\beta=(\pi^2/15)((k_{B}T_{0})^{4}/(\epsilon_0 \hbar^3 c^3))(k^{\mathrm{s}}/\nu_{0})=k^{\mathrm{s}} E_0^2/\nu_{0}$
is the clock frequency, $\alpha$ the fine-structure constant and $E_0\simeq 831.9$~V.m$^{-1}$ the rms value of the BBR electric field at $T_0$. For Rb, an experimental determination of the Stark shift coefficient $k_{\mathrm{Rb}}$ is obtained based on the measurement of the ratio of the Stark coefficients for Rb and Cs, $k_{\mathrm{Rb}}/k_{\mathrm{Cs}}$~=~0.546(5) \cite{Mowat1972}, combined with the experimental value of the scalar Stark coefficient of Cs, $k_{\mathrm{Cs}}^{\mathrm{s}}$~=~$-2.282(4)\times 10^{-10} \mathrm{Hz}/(\mathrm{V}.\mathrm{m}^{-1})^{2}$  \cite{Guena2012}\cite{Rosenbusch2007}\cite{Simon1998}. This approach gives $\beta _{\mathrm{Rb}}^{\mathrm{exp}}=-1.2727 \times 10^{-14}$ with an uncertainty of $1.1\times 10^{-2}$ including an uncertainty for the lack of knowledge of the Rb tensor contribution to the Stark shift measurement of \cite{Mowat1972}. $\beta _{\mathrm{Rb}}^{\mathrm{theo}}$ and $\beta _{\mathrm{Rb}}^{\mathrm{exp}}$ differ by about one standard uncertainty, which is not statistically significant.
For the BBR shift correction in {\forb}, we use this second experimental determination, and the theoretical estimation of $\epsilon_{\mathrm{Rb}}$ with a 10 percent uncertainty. At 300~K, this leads to an uncertainty of $1.35\times 10^{-16}$ associated with the atomic parameters.

In FO2, the innermost layer surrounding the interrogation region is made of an aluminium alloy and is well isolated from the environment to ensure good temperature uniformity of the atom's environment. The temperature of the vacuum chamber smoothly follows the room temperature with a time constant of approximately one day and no large gradient exists between the inner region and the environment.
This essentially removes concerns about the effect of the necessary openings (for the atomic trajectories,
pumping, etc.) on the effective blackbody temperature seen by the atoms. Fig. \ref{fig_Tstability} shows the long term instability of these temperatures.
To control the temperature and thermal gradients in the interrogation, we use three main platinum resistors evenly distributed along the innermost aluminum layer of the vacuum chamber. The three sensors were initially compared to a reference sensor to within 0.01~K, whose calibration was verified by the LNE temperature section. This allows us to verify that the temperature is uniform in the interrogation region to better than 0.02~K. During operation of {\forb} as a frequency standard, we monitor the three temperatures every  hour and take the time-averaged value from the central probe as the blackbody radiation temperature for the corresponding clock measurement. We take a final temperature uncertainty of 0.2~K corresponding to the uncertainty of the sensors originally specified by the manufacturer. All uncertainties are combined quadratically to give the overall BBR uncertainty of $1.4\times 10^{-16}$.

\subsection{Atom number dependent shift}\label{subsec_Colpul}

This shift has contributions from collisions and from cavity pulling \cite{Bize2001}.
At the same atomic density, the shift for Rb is typically more than 30 times smaller than for Cs \cite{Sortais2000}\cite{Fertig2000} when no shift cancellation method is used \cite{Szymaniec2011}. In {\forb}, the shift is $\leq 1\times 10^{-15}$ at the highest atom number.
We measure the shift during clock operation by switching between full and half atom number (HD/LD configuration for high/low atomic density) every 50 cycles. The atom number is changed via the microwave power in the state-selection cavity. With this method, the cloud distributions at high and low atom number differ, which can bias the extrapolation to zero atomic density. Given the duration of the microwave pulse (2~ms), the cloud velocity and the shape of the cavity field, we estimate the dispersion of the Rabi frequency over the size of the vertically moving cloud to be less than $\pm 6\%$. The corresponding maximum possible error in the determination of the collision shift coefficient is $\pm 19\%$. This, in turn, translates into an error in the extrapolated frequency equal to $14\%$ of the observed shift at high density and a typical fractional frequency uncertainty of $1.5\times 10^{-16}$.  The real-time HD/LD measurement of the shift has also a Type~A contribution which depends on the measurement duration, typically $1.4\times 10^{-16}$ for 15 days.

It is slightly paradoxical that the collision shift is the leading term in the uncertainty budget of Table \ref{tab_accuracy}, given the small size of the effect in $^{87}$Rb.
This situation ended up occurring in our system because the interrupted adiabatic passage method that we use for Cs \cite{Pereira2002} was not straightforward to implement in the context of the dual fountain configuration. In future, we could reduce the uncertainty of the atom number dependent shift using the method proposed in \cite{Gibble2012}, still using the microwave interaction in the state-selection cavity.

Note that in FO2, the two Rb and Cs clouds are launched almost simultaneously but at slightly different velocities. Thereby, interspecies collisions are avoided during the interrogation period \cite{Guena2010}.

\subsection{Distributed cavity phase shift} \label{subsec_DCP}

The distributed cavity phase (DCP) shift is a residual first order Doppler shift associated with the spatial phase distribution of the microwave field inside the Ramsey cavity. 
We evaluate this shift by applying the approach established in \cite{Guena2011} for {\focs}, which is based on a theoretical model of the cavity field developed in \cite{Li2004}\cite{Li2010b}.
In this approach, the leading contributions from the lowest azimuthal terms $m = 0$, $m = 1$ and $m=2$ of the phase distribution in the cavity are considered. Table I in \cite{Guena2012} gives the details of these contributions for {\forb} specifically, as well as all other LNE-SYRTE fountains. The $m = 1$ term, associated with an effective tilt of the launch direction along the microwave feed axis, is measured by differential measurements with the cavity fed asymmetrically from one or the other side. From such measurements we estimate an uncertainty of $\sim 0.2~$mrad in the effective tilt parallel to the microwave feed axis. We estimate an uncertainty of $0.7~$mrad in the perpendicular direction based on the maximum detected atom number as a function of tilt.
The tilt sensitivities measured in the nominal symmetric feeding configuration are $(1.8\pm 1.0)\times 10^{-16}$~mrad$^{-1}$ and $\leq 0.9\times 10^{-16}$~mrad$^{-1}$ for parallel and perpendicular tilts respectively. The even $m=0$ and $m=2$ terms are calculated from the phase distributions for the {\forb} cavity used as inputs into simulations of the atom travel paths. The overall correction for the DCP shift is $(0.4 \pm 1.0)\times 10^{-16}$.

\subsection{Microwave lensing shift}

The microwave lensing effect \cite{Gibble2006} leads to a frequency shift which was first calculated in a detailed manner in \cite{Li2011} and \cite{Weyers2012}. We have applied this same approach to the LNE-SYRTE fountains. Proper calculation of the shift requires taking into account the actual experimental parameters such as the shape of the Ramsey cavity field, the space and velocity distributions of the atomic cloud and its truncation by the cavity apertures, as well as the detection inhomogeneities and the contrast of Ramsey fringes. The results of these calculations were reported in Table II of \cite{Guena2012}. The overall correction for {\forb}~is $(-0.7\pm 0.7)\times 10^{-16}$.

\subsection{Microwave leakage and spectral purity}\label{subsec_leakage}

The architecture of the microwave synthesizer of {\forb} is schematized in Fig. \ref{fig_metchain}.
To evaluate microwave leakage shifts, this specially developed synthesizer is equipped with a RF switch based on a Mach-Zehnder interferometer design which is free of phase transients \cite{Santarelli2009}. In the switched configuration of the fountain, the microwave signal is switched on and off when the atomic cloud is in the cavity cut-off waveguides just before and after each passage into the Ramsey cavity.
Checks for leakages by differential frequency measurements, switched $\it{vs}$ continuous microwave signal, as a function of microwave power \cite{Santarelli2009} put an upper limit to a putative frequency shift due to microwave leakage at $0.5\times 10^{-16}$ for operation with the nominal microwave power. The test is repeated from time to time at the nominal power.
We also occasionally (once a year) analyze the interrogation signal using a dedicated triggered phase transient analyzer with microradian resolution \cite{Santarelli2009}. Both phase transients associated with the switching of the microwave and with phase perturbations synchronous with the fountain cycle can be addressed. No spurious effect is detected at sensitivities as low as
 1~$\mu$rad.s$^{-1}$, corresponding to fractional frequency resolution of $2.3\times 10^{-17}$.

Concerning spectral purity, no spurious sidebands are observable above -70~dB of the 6.834~GHz carrier down to a resolution bandwidth of a few Hz. We note that the cycle period (1.6045 s) is chosen so as not to be synchronous with the 50 Hz line (and first harmonics) period.

We establish a global uncertainty $<0.5\times 10^{-16}$ \cite{Santarelli2009}.

\subsection{Rabi and Ramsey pulling shifts}

The Rabi pulling shift depends on residual populations $P_{m_{F}}$
of the $m_{F}\neq 0$ Zeeman sub-levels relative to that of the $m_{F} = 0$ initial clock state (see, for instance, \cite{Gerginov2010}). We measured these populations and found $P_{\pm 1}\approx 10^{-3}$ for the leading ones, with a difference less than $10^{-4}$. Using \cite{Cutler1991} we estimate an upper limit for Rabi pulling of $10^{-20}$.
Similarly, we measured the $0 \rightarrow \pm 1$ $\sigma^{\pm}$ transition probabilities relative to the clock transition probability on resonance: we found $1.5\times 10^{-2}$ with an asymmetry (defined as $(\sigma^{+}- \sigma^{-})/(\sigma^{+}+ \sigma^{-})$) that is less than $3\times 10^{-3}$. With some estimations regarding the transverse component of the microwave field and coupling of the atom to this field, using again \cite{Cutler1991} we infer an upper limit of the Ramsey pulling shift less than $10^{-19}$.

We note here the favorable properties of Rb when comparing to Cs. The Zeeman splitting
is larger by a factor of 2 for Rb. The duration of the passage in the microwave cavity is larger
by 1.4 for Rb, and correspondingly the Rabi frequency for a $\pi/2$ pulse area is smaller for Rb. In
addition, since the Rb cavity has larger dimensions, the transverse microwave fields are
comparatively smaller over the cloud dimensions. These properties reduce the Rabi and Ramsey pulling shifts by a
factor of more than 10.

\subsection{Background gas collisions}

This effect comes from collisions between cold atoms and residual background gases in the interrogation region. The approach generally used so far in Cs fountains to estimate this shift was relying on pressure shifts measured in vapor cells near room temperature. The physical basis for this approach is poorly justified given that conditions found in vapor cells are considerably different than those encountered in atomic fountains. Recently, a model of background collision shift adapted to atomic fountains was developed \cite{Gibble2013}. In FO2, the use of 2DMOT sources (see Fig.~\ref{fig_FontaineRbCs}) leads to a negligible alkaline background vapor in the interrogation region. Thus, a reasonable assumption is that the background collision shift is predominantly due to H$_2$. The model of Ref.~\cite{Gibble2013} applied to Rb-H$_2$ collisions predicts a fractional frequency shift of $\sim 10^{-17}$ for a background pressure of $10^{-9}$~mbar. Using ion pump currents carefully corrected for electronic offsets, we estimate the residual pressure in the interrogation region to be $2.7\times 10^{-9}$~mbar, which yields a shift of $<3\times 10^{-17}$. Reference \cite{Gibble2013} also relates the background collision shift to the fractional loss of cold atoms during interrogation, predicting that the shift is less than $3\times 10^{-17}$ if no more than 20\% of the cold atoms are lost. We have measured (simultaneously for Cs and Rb) the loss of atoms as a function of pressure and used these measurements to determine that the lost fraction under nominal conditions is about 16\% (both for Cs and Rb). According to the second prediction, this also yields a shift of $<3\times 10^{-17}$. Given the significant uncertainty associated with  estimating the pressure in the interrogation region based on ion pump currents (a factor of 2 is possible), given the modest range of atom loss that could be explored in the second test (20\%) and given the lack of experimental confirmation of the model of Ref.~\cite{Gibble2013}, we consider that an upper bound of $10^{-16}$ is a reasonable estimation of the uncertainty.

\subsection{Light shifts}

The dual fountain configuration of FO2 allows the usual approach: All laser beams are blocked with mechanical shutters out of their application periods and most importantly during the Ramsey interrogation. The proper functioning of shutters is checked from time to time, as well as the absence of stray light.

\subsection{Relativistic effects}\label{subsec_redshift}

Due to the motion of the atomic cloud and to gravity, relativistic effects must be taken into account. During the ballistic flight above the Ramsey cavity, the proper time of the atoms $\tau$ differs from the proper time $t$ for the microwave cavity field. According to General Relativity, we have $d\tau\simeq\left(1+\left[g.z(t)-v^2(t)/2 \right]/c^2\right)dt$, where $g$ is the local acceleration due to gravity, $c$ is the speed of light, $z(t)$ is the height of the atom above the cavity at time $t$ and $v(t)$ is the velocity of the atom. From this equation, we find that the reference point where the unperturbed atomic frequency is realized, is located $h/3$ above the cavity center~\footnote{The phase of the atomic coherence cumulated during the Ramsey time is $\Delta\phi_{at}=\int \omega_{at} d\tau$. The phase cumulated by the field is $\Delta\phi=\int \omega dt$. The lock to the atomic transition has the effect of tuning $\omega$ so that the condition $\Delta\phi_{at}=\Delta\phi$ is realized. This gives $\omega=\omega_{at}(1+g h /3 c^2)$.}.
$h$ denotes the height of the apogee above the cavity. We have $h/3=0.146$~m in {\forb} and the reference point is shown in Fig.~\ref{fig_FontaineRbCs}. Due to the atomic position and velocity distributions and to the distribution of the microwave field, there are small deviations from the nominal trajectory of the cloud center used to determine the reference point. This results in a frequency uncertainty well below $5\times 10^{-18}$ (equivalent to $5$~cm) from residual unmodelled relativistic effects.

When the {\forb} SFS is compared to remote clocks, relativistic effects between these remote clocks and the above defined reference point must be taken into account. The particular case of linking {\forb} to TAI is discussed in Sect.~\ref{subsec_Links}.

\section{Frequency metrology chain}\label{sec_MetChain}

\begin{figure}[h]
\hspace{0mm}
\includegraphics[angle=0,width=\linewidth]{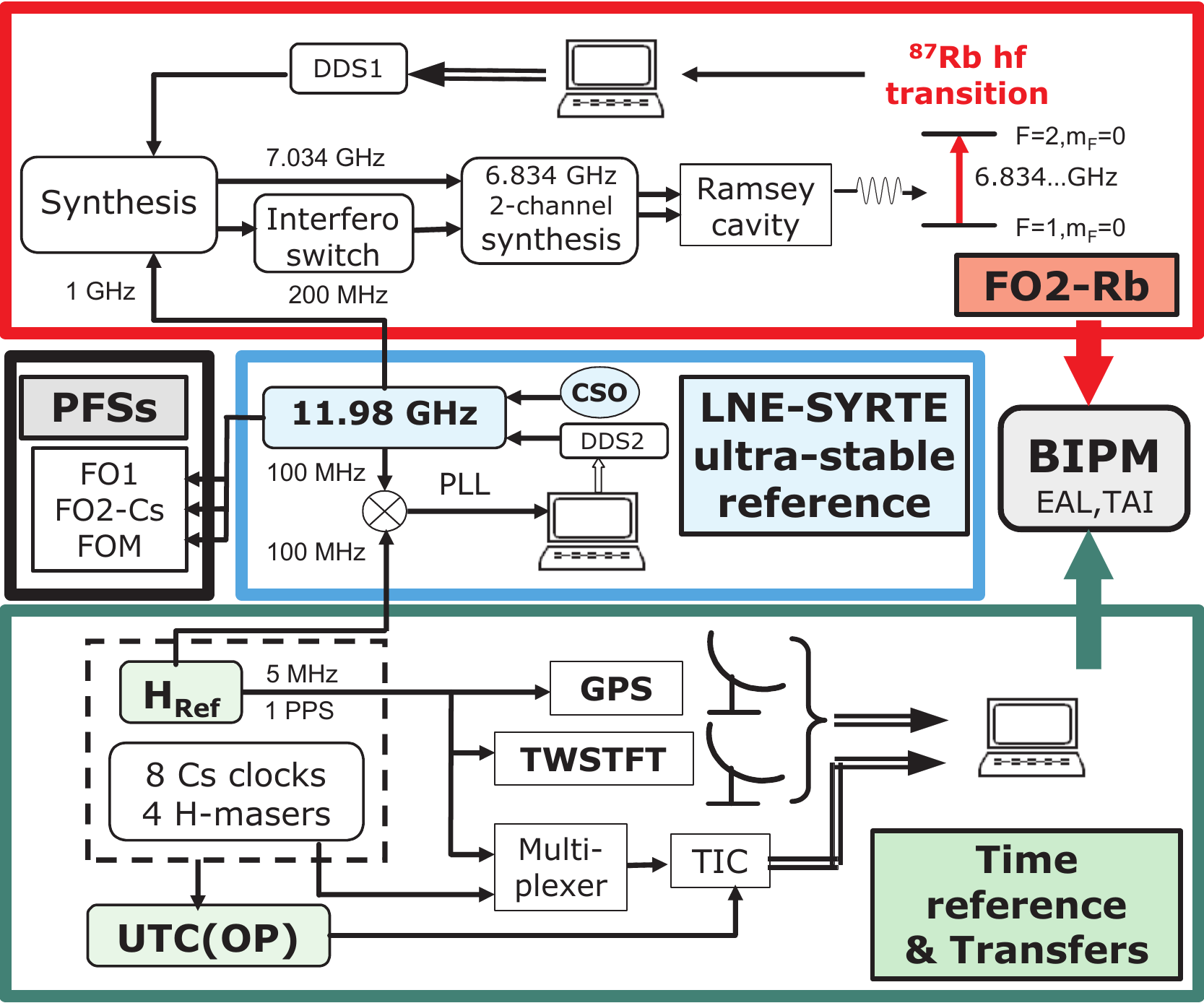}
 \caption{{\forb} frequency chain and connection to international time scales. DDS: direct digital synthesizer; CSO: cryogenic sapphire oscillator from the University of Western Australia; PLL: digital phase lock loop; PFSs: LNE-SYRTE Cs fountain Primary Frequency Standards; H$_{\mathrm{Ref}}$: LNE-SYRTE reference hydrogen maser; UTC(OP): real time prediction of the Coordinated Universal Time (UTC) for France; TIC: time interval counter. \label{fig_metchain}}
 \end{figure}

The secondary representation of the second under consideration in this work is the $^{87}$Rb ground state hyperfine transition.
In this section, we describe the metrological chain between the realization of the frequency of this transition in {\forb} and the international atomic time. This architecture is schematized in Fig. \ref{fig_metchain}.

\subsection{LNE-SYRTE ultra-stable reference} \label{subsec_synthesis}

At the heart of the frequency distribution scheme at LNE-SYRTE is an ultra-stable reference signal derived from a cryogenic sapphire oscillator (CSO) developed at the University of Western Australia \cite{Luiten1995b}\cite{Tobar2006}. This oscillator is left free-running, delivering a 11.932~GHz signal with low phase noise. As represented in Fig. \ref{fig_metchain} (middle section, blue), a frequency offset stage \cite{Chambon2005} shifts the CSO signal to 11.98~GHz whose tunability is provided by a computer-controlled direct digital synthesizer (DDS2). The 11.98~GHz signal reproduces the excellent spectral purity of the CSO. This highly stable signal is down-converted to 1~GHz and to 100~MHz \cite{Chambon2007}. The phase of this 100~MHz signal is compared to the phase of the 100~MHz output of a reference hydrogen maser (H$_{\mathrm{Ref}}$). A digital phase-lock loop with a time constant of $\sim 1000$~s is implemented by acting on DDS2. The 11.98~GHz, 1~GHz and 100~MHz signals are thereby made phase-coherent with the oscillation of the maser, combining the excellent short-term stability of the CSO with the mid- and long-term stability of H$_{\mathrm{Ref}}$ as well as the connection to local time scales. As sketched in Fig. \ref{fig_metchain} (middle section), the same ultra-stable references are delivered to LNE-SYRTE fountain PFSs.

\subsection{{\forb} microwave synthesizer}

A low phase noise synthesizer is used to convert the 1~GHz ultra-stable reference signal to the $^{87}$Rb hyperfine frequency. This synthesizer is described in detail in \cite{Guena2010}. Figure \ref{fig_metchain} (upper section, red) highlights the main features. The 1~GHz signal is up-converted to a 7.034~GHz signal, tunable with microHertz resolution thanks to a computer-controlled DDS (DDS1). The 1~GHz is also divided to generate 200~MHz which is passed through a Mach-Zehnder interferometer switch \cite{Santarelli2009} already mentioned in Sect. \ref{subsec_leakage}.
The 7.034~GHz and 200~MHz signals are then mixed in a 2-channel device generating two signals at the $^{87}$Rb hyperfine frequency of $6.834$~GHz
feeding the two inputs of the {\forb} Ramsey cavity. Power and phase adjustments of the 7.034~GHz on one of the channels are used to balance the cavity feeds in order to minimize the $m=1$ term in the DCP shift (see Sect. \ref{subsec_DCP}). Frequency corrections applied to DDS1 are the basis for determining frequency stabilities, for investigating systematic shifts and for measuring the frequency of the reference signal with respect to the {\forb} standard.

The phase noise power spectral density, the long term phase stability and the putative phase transient in this synthesizer were extensively characterized following approaches developed for Cs \cite{Santarelli2009}\cite{Chambon2007}. The phase noise of the synthesizer does not limit the short term stability achievable in {\forb} down to $<10^{-14} \tau^{-1/2}$. This allows the {\forb} fountain to operate at the quantum projection noise limit with a stability of $3.0$ parts in $10^{14}$ at $1$~s for the presently achievable highest atom number.

\subsection{Link to local and international time scales} \label{subsec_Links}

The previously mentioned reference hydrogen maser H$_{\mathrm{Ref}}$ is also the input of the GNSS and TWSTFT satellite links, as shown in the lower section of Fig. \ref{fig_metchain} (green). H$_{\mathrm{Ref}}$ is one of LNE-SYRTE's commercial clocks participating in the elaboration of TAI by the BIPM.
One of these clocks (H$_{\mathrm{Ref}}$ since October 2012) is used to feed a phase micro stepper to generate UTC(OP), the real time realization of the prediction for France of the Coordinated Universal Time (UTC). All commercial clocks are alternately compared to UTC(OP) using a switch unit and the same 1~PPS time interval counter. Every month, time differences between each clock (including H$_{\mathrm{Ref}}$) and UTC(OP) sampled along the internationally agreed 5-day grid (Modified Julian Date MJD ending with 4 or 9), are sent to the BIPM. Data from \emph{Circular~T} are used to steer UTC(OP) to UTC. Since October 2012, fountain PFSs data are also used to control the frequency of UTC(OP) on a daily basis resulting in a significant improvement of the stability of this time scale.

When contributing to TAI, relativistic gravitational red shift between the reference point of {\forb} SFS (see Sect.~\ref{subsec_redshift}) and the rotating geoid must be taken into account. This correction amounts to $(-65.4\pm 1)\times 10^{-16}$ corresponding to a 1~m uncertainty in height. A leveling campaign at Observatoire de Paris will soon allow us to improve the determination of this correction.

\subsection{Data acquisition and processing} \label{subsec_DataTreatment}

The {\forb} fountain is operated in a frequency standard mode, providing at each fountain cycle a value for the ratio between the frequency of the $^{87}$Rb hyperfine transition and the frequency of the $100$~MHz output of the maser H$_{\mathrm{Ref}}$.

The atom number dependent shift (see Sect. \ref{subsec_Colpul}) is continuously measured by running an interleaved sequence of 50 cycles at a high atom number $n_\mathrm{H}$ and 50 cycles at a low atom number adjusted to $n_\mathrm{L}\simeq 0.5 \times n_\mathrm{H}$. Over a typical period of 10~days defining a fountain run, the shift per atom is determined based on DDS1 (Fig. \ref{fig_metchain}) mean frequency corrections and on the mean detected atom numbers $\langle n_\mathrm{H}\rangle$ and $\langle n_\mathrm{L}\rangle$. This mean shift per atom is used to correct cycle-to-cycle data for the atom number dependent shift. All other systematic corrections as well as the gravitational red shift correction, are then applied. Second order Zeeman and BBR corrections are based on the mean values of the static magnetic field and temperature monitored as described in Sect. \ref{sec_accuracy}. This finally produces a post-processed series of cycle-to-cycle measurements of H$_{\mathrm{Ref}}$ with the {\forb} frequency standard.

A second layer of data processing is applied for comparisons to primary frequency standards at LNE-SYRTE. The frequency is averaged over intervals of 0.01 and 0.1~day, synchronous for all standards. The frequency differences ({\forb} - PFS) are then computed, removing the common-mode frequency of the hydrogen maser H$_{\mathrm{Ref}}$. These comparisons are used to estimate the stability between standards (Sect.  \ref{sec_measurements}) and to determine the absolute frequency of the $^{87}$Rb hyperfine transition (Sect. \ref{sec_hfs measurements}).

Every month, calibrating TAI requires estimations of the frequency of H$_{\mathrm{Ref}}$ averaged over periods along the internationally agreed 5-day grid. In order to provide such estimations based on the $^{87}$Rb SRS, we start from the measurements of H$_{\mathrm{Ref}}$ with {\forb} averaged over 0.2 day. In order to deal with dead times, data are fitted to a straight line yielding the average frequency of H$_{\mathrm{Ref}}$ for the period. The Type~A uncertainty $u_{\mathrm{A}}$ of this determination is estimated based on the fractional frequency instability of the residuals observed after subtracting the linear drift of the maser. Typically, $u_{\mathrm{A}}\simeq 2 \times 10^{-16}$ over a period of 20 to 30 days. Another uncertainty $u_{\mathrm{l/Lab}}$ accounts for the impact of dead times in the measurement of H$_{\mathrm{Ref}}$ and for possible phase fluctuations in the connecting cables. The overall amount of dead time is generally less than 10-20$\%$ and the overall link uncertainty $u_{\mathrm{l/Lab}}$ is typically $10^{-16}$. The approach for determining $u_{\mathrm{l/Lab}}$ is identical to the one used for our PFSs (see our calibration reports published by the BIPM).

In order to handle the metrological chain in Fig.~\ref{fig_metchain}, we developed a software infrastructure automating several important tasks. On an hourly basis, {\forb} data as well as PFSs data are collected and backed-up. The above described generation of fountain data corrected for all systematic frequency shifts is performed at a preliminary level. A graphical interface displays these data, the fractional frequency instability of ({\forb} - H$_{\mathrm{Ref}}$), as well as other important parameters, such as atom numbers, transition probabilities, collision frequency shift, static magnetic field, temperatures, etc. A similar approach is used for the ultra-stable reference based on the cryogenic sapphire oscillator and for other subsystems. With this, an operator can have an overview of the status of the whole metrological chain in only a few minutes. On a daily basis, these data are further processed to provide H$_{\mathrm{Ref}}$ calibrations and comparisons between fountains, providing a further mean to assess the whole system. This software infrastructure in fact automatically generates data ready for calibrations of time scales or other frequency measurements. Yet, these data are critically scrutinized and re-processed whenever necessary for final applications.

An additional functionality of the software infrastructure is to perform real-time detection of anomalies such as phase/frequency jumps of the ultra-stable reference signal or interruptions of fountain operation due to laser unlocks. Email alerts are sent to operators of the system. Based on these anomalies, periods of time over which fountain data are automatically discarded are defined. The softwares also deal with the longer breaks due, for instance, to user interruptions of the fountains for small optimizations (coupling of laser light into optical fibers, etc.) typically done every 1 to 2 weeks. Another cause of interruption in our system are the liquid helium refills of the cryogenic sapphire oscillator done every 26~days.

\section{Measuring with the {\forb} frequency standard}\label{sec_measurements}

\begin{figure}
\includegraphics[angle=0,width=\linewidth]{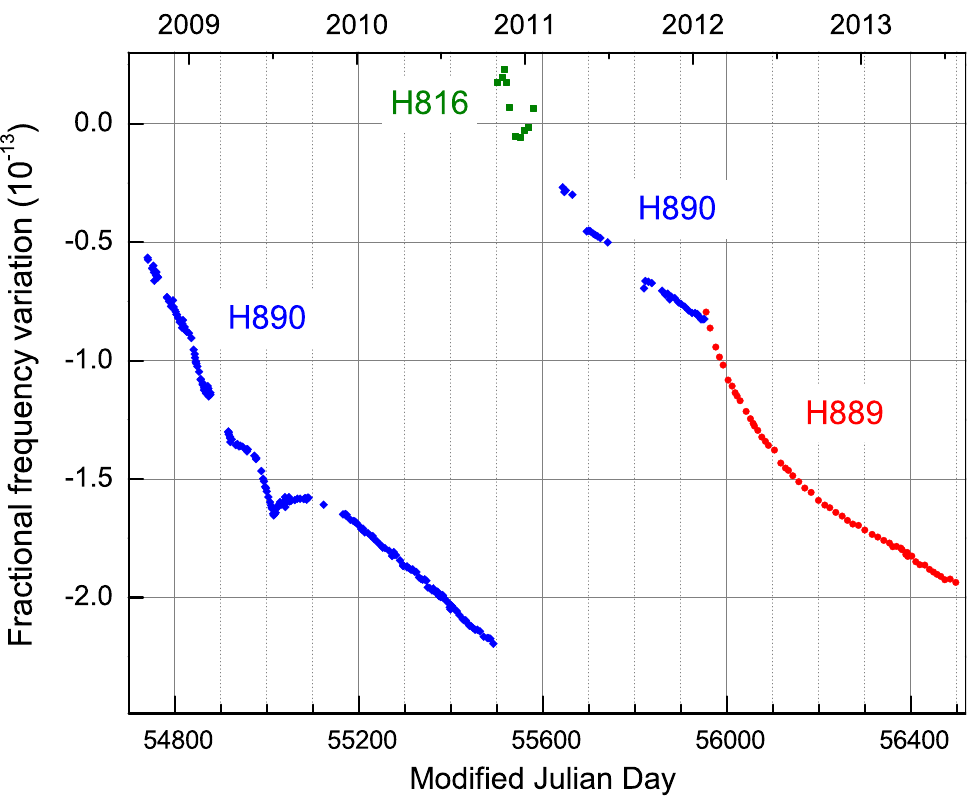}
\caption{Frequency variations of the LNE-SYRTE reference hydrogen maser as measured with the {\forb} Secondary Frequency Standard from October 2008 to end of July 2013. HXXX is H-maser 1400XXX in the BIPM clock nomenclature
$^{a}$. Gaps in the data correspond to interruptions of {\forb} for studies of systematic effects, or refurbishing the set-up.
\newline
$^{a}${H889, H890 are model CH1-75A purchased to IEM Kvarz (Russia), H816 is model PAR-2001 purchased to Sigma Tau Standards Corporation (USA) currently Symmetricom, Inc. (USA).}}
\label{Fig_MaserFO2Rb}
\end{figure}

\begin{figure}
\includegraphics*[width=\linewidth]{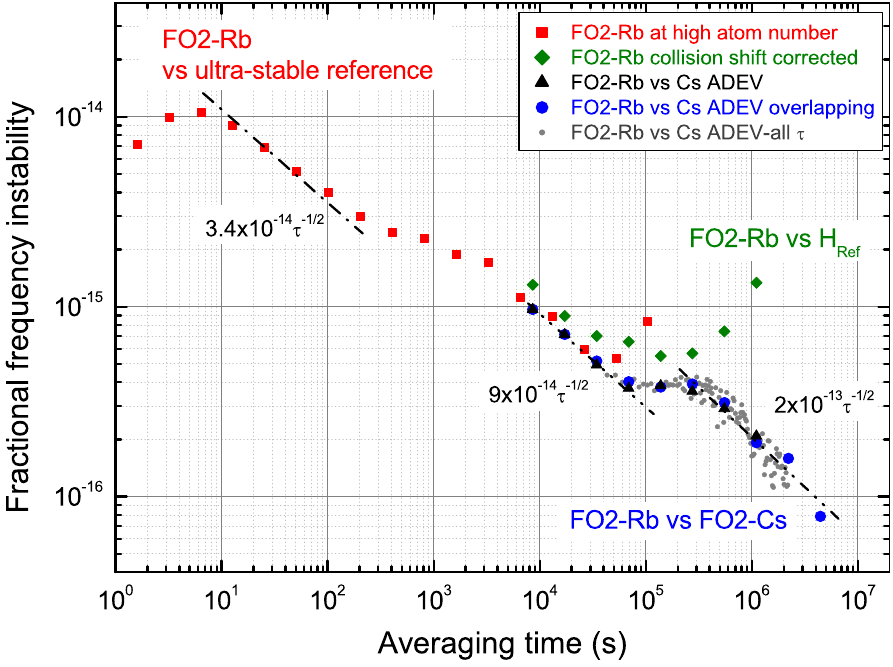}
\caption{Fractional frequency instability characterized in terms of the Allan standard deviation observed with the {\forb} Secondary Frequency Standard and the metrology chain of Fig.~\ref{fig_metchain}. Red squares: Instability obtained with {\forb} operated at high atom number. The first points reflect the dynamics of the digital servo system locking the $6.8$~GHz microwave signal to the atomic transition. For $\tau > 1000$~s, the phase lock loop of the ultra-stable reference to the H-maser is effective. Therefore, data reflects the behavior of the H-maser (here, data sample taken with H889). Green diamonds: Instability obtained with {\forb} operated as a frequency standard, \emph{i.e.} including both high and low atom number data and correcting for all systematic frequency shifts. These data represent measurement of the H890 hydrogen maser frequency over six months with the {\forb} secondary frequency standard. The drift of the H-maser becomes visible above 10$^{5}$~s.
Black triangles: Instability of the comparison between {\forb} and {\focs} over the same six months. This measurement is obtained thanks to the simultaneous operation of the two standards allowing for the elimination of the common-mode H-maser frequency over synchronous intervals of 0.1~day (see text). Grey small dots: Allan standard deviation computed for all $\tau$. Blue large dots: overlapping Allan standard deviation. The shoulder at $\sim 3\times 10^5$~s corresponds to the duration of $\sim 10$~days over which corrections for the atom number dependent shift are applied (see Sect. \ref{sec_measurements}). The dash-dotted lines are guides for the eye with slopes $3.4\times 10^{-14}\tau^{-1/2}$, $9\times 10^{-14}\tau^{-1/2}$ and $2\times 10^{-13}\tau^{-1/2}$.
\label{Fig_RbCsAvar}}
\end{figure}

Figure \ref{Fig_MaserFO2Rb} shows more than one thousand days of measurements of the frequency of our reference hydrogen maser H$_{\mathrm{Ref}}$ with {\forb} using the metrological chain of Fig. \ref{fig_metchain}. Over the years, three different H-masers were involved.
Figure \ref{Fig_RbCsAvar} displays frequency instabilities associated with these measurements. Red squares show the short term instability of 3.4$\times10^{-14}$ at 1~s obtained at high atom number over a two week long sample.
Such a short term instability is enabled by the use of the ultra-stable signal derived from the CSO (Fig. \ref{fig_metchain}). The behavior of the 3 first points reflects the lock of the {\forb} synthesizer to the central Ramsey fringe.  The shoulder at $\sim 1000$~s comes from the lock of the ultra-stable reference signal to the H$_{\mathrm{Ref}}$ maser. The instability beyond 1000~s is determined by H$_{\mathrm{Ref}}$. Green diamonds show the instability of H$_{\mathrm{Ref}}$ measured by {\forb} when both high and low atom number data are used, with corrections for all systematics applied as described in Sect. \ref{subsec_DataTreatment}. The long term drift of the H-maser ($\sim -1.5 \times10^{-16}\mathrm{d}^{-1}$) is clearly seen after a few days.

Black triangles in Fig. \ref{Fig_RbCsAvar} display the fractional frequency instability for a comparison between {\forb} and {\focs}. This plot shows a change of slope, from $9\times 10^{-14}\tau^{-1/2}$ for $\tau < 10^{5}$~s to $2\times 10^{-13}\tau^{-1/2}$ for $\tau > 4\times 10^{5}$~s ($\sim 5$~days). This behavior is due to the method used to measure and correct for the atom number dependent shift as described in Sect. \ref{subsec_DataTreatment}. The averaging time at which the change of slope is seen corresponds to the $\sim 10$~day period over which the shift per atom is determined. This behavior is well-reproduced by simulation of the data and it is present both in {\forb} and {\focs}. A fractional frequency instability of $\mathrm{1.4}\times 10^{-16}$ is reached at $2\times 10^6$~s. Grey dots show the fractional frequency instability computed for all $\tau$, and blue dots show the overlapping Allan deviation.

Regarding our method to determine the atom number dependent shift, our parameters ($n_\mathrm{L}\simeq 0.5\times n_\mathrm{H}$, 50 cycles at both $n_\mathrm{H}$ and $n_\mathrm{L}$) were kept so far for various historical reasons associated with the softwares used to run the fountains and process the data. We note however that they are not optimum for the long term stability \cite{Szymaniec2011}. In the future, optimization of our parameters could reduce the measurement time by up to a factor of 2 for a given resolution.

\section{Absolute frequency measurements of the $^{87}$Rb ground state hyperfine transition}\label{sec_hfs measurements}

 \begin{figure}[h]
 \includegraphics[width=\linewidth]{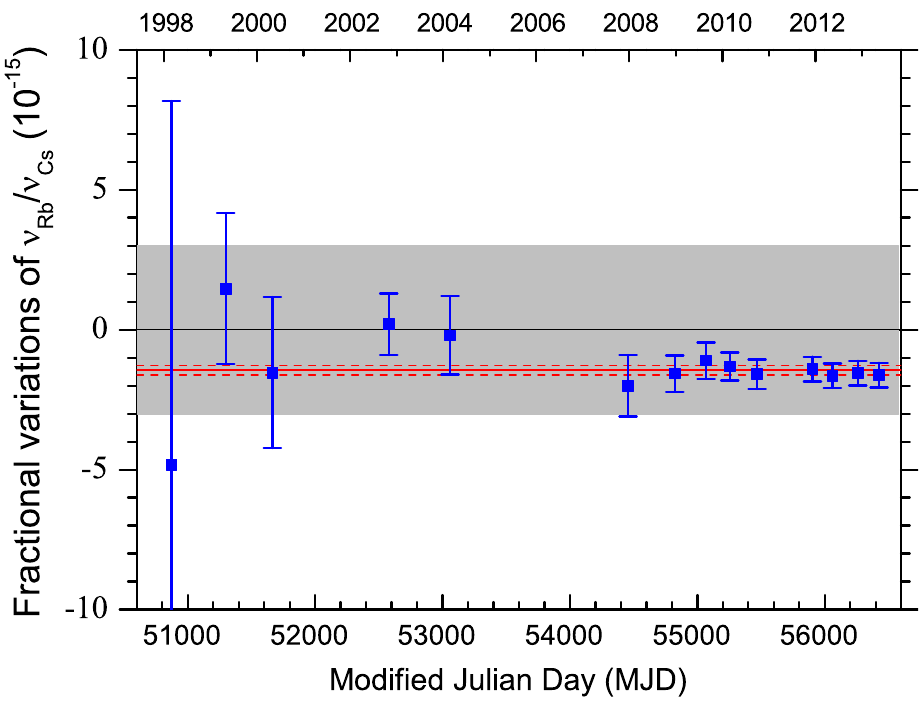}
 \caption{Measurements of the $\nu_{\mathrm{Rb}}/\nu_{\mathrm{Cs}}$ hyperfine frequency ratio measured with {\forb} and the LNE-SYRTE PFSs. The graph shows fractional variations of the ratio with respect of the value defined by the 2004 recommended value for $\nu_{\mathrm{Rb}}$: 6 834 682 610.904 324~Hz \cite{CCTF2004}\cite{CIPM2007}. The error bars are the total 1~$\sigma$ uncertainties, here dominated by Type~B uncertainties.
 The red lines show the result of the fit of the data points to a constant with inverse quadratic weighting. This fit is used to test the consistency of the data (see text). The grey area displays the recommended uncertainty of the 2004 recommended value: $3 \times 10^{-15}$.
 Table \ref{tab_CompRbCs} gives the conditions of the comparison for each point of the graph.}
 \label{Fig_RbCsFinal}
 \end{figure}

 Several measurements of the ratio $\nu_{\mathrm{Rb}}/\nu_{\mathrm{Cs}}$ between the ground-state hyperfine frequencies of $^{87}$Rb and $^{133}$Cs were performed using {\forb} and the LNE-SYRTE fountain ensemble (Fig. \ref{fig_metchain}), with a method similar to the one discussed in Sect. \ref{sec_measurements} \cite{Bize1999}\cite{Marion2005}\cite{Guena2010}\cite{Guena2012a}. These measurements directly yield determinations of the
absolute frequency of the $^{87}$Rb ground-state hyperfine transition in the SI system. Figure~\ref{Fig_RbCsFinal} summarizes these measurements. The origin of the vertical axis corresponds to the value of $\nu_{\mathrm{Rb}}$ recommended in 2004 \cite{CCTF2004}\cite{CIPM2007}. Error bars in this plot display the overall standard uncertainty (1~$\sigma$), which combines the Type~B uncertainties of both {\forb} and the Cs fountains as well as the statistical uncertainty of the comparison. The quadratic sum of these 3 contributions is dominated by Type~B uncertainties.

Table \ref{tab_CompRbCs} indicates the middle date of the comparison for each data point in Fig. \ref{Fig_RbCsFinal}, the primary reference(s) for the measurement, and the operation mode of {\forb}. The duration of comparisons varies from a few weeks for early measurements to several months. Measurements were quasi-continuous over the last three years. When data from two or three Cs references were available, we computed a weighted average taking into account the total uncertainty and the duration for each Cs/Rb pair.

\begin{table}
\begin{center}
\caption{Conditions of the measurements displayed in Fig. \ref{Fig_RbCsFinal}}\label{tab_CompRbCs}

\begin{tabular}{cccc}
\hline \hline
$\sharp$ & Middle date  & PFS  & FO2 mode \\
\hline 
1 &  50873  & FO1  & Rb  \\
2 &  51299  & FO1, FOM  & Alt Rb/Cs \\
3 &  51665  & FO1, FOM  & Alt Rb/Cs \\
4 &  52579  & FOM  & Alt Rb/Cs \\
5 &  53060  & FOM  & Alt Rb/Cs \\
6 &  54458  & FOM  & Alt Rb/Cs \\
7 &  54824 & FO1, FO2 & Dual Rb/Cs \\
8 &  55069 & FO1, FO2 & Dual Rb/Cs \\
9 &  55255 & FO1, FO2 & Dual Rb/Cs \\
10 & 55469 & FO1, FO2, FOM & Dual Rb/Cs \\
11 & 55895 & FO1, FO2, FOM & Dual Rb/Cs \\
12 & 56060 & FO2, FOM & Dual Rb/Cs \\
13 & 56261 & FO2, FO1 & Dual Rb/Cs \\
14 & 56425 & FO2 & Dual Rb/Cs \\
\hline \hline
\end{tabular}
\end{center}
{The first column is the data point number, the second is the middle date of the measurement and the third is the Cs PFS reference(s). The last column specifies the mode of operation of the FO2 fountain. Early on, FO2 was operated with Rb only. Then, it was operated either with Rb or with Cs. Most recently, FO2 was operated with Rb and Cs simultaneously \cite{Guena2010}.}
\end{table}

The consistency of the data can be tested by a weighted linear least square fit to a constant with inverse quadratic weighting, \textit{i.e.} weights inversely proportional to the square of error bars. The result of this fit gives $-14.4\times 10^{-16}$ with a standard error of $1.7\times 10^{-16}$. It is shown by the solid and dashed red lines in Fig.~\ref{Fig_RbCsFinal}. The maximum deviation from the fitted value observed in the data is 1.47~$\sigma$ (data point $\sharp 4$). The reduced chi-square is $\chi^{2}$ = 0.41, and the goodness-of-fit $Q = 0.97$ \cite{NR1992}. The Birge ratio \cite{Birge1932}, \textit{i.e.} the square-root of the reduced chi-square, is $R_{\mathrm{B}}=0.64$. All three indicators show that the data set is consistent. In fact, they indicate a spread of data smaller than expected for the normal probability distribution. This is likely due to the fact that some systematic biases are correlated over time despite the large number of modifications made over the years. Possibly, this could also be partly due to the fact that some Type~B uncertainties are slightly overestimated. For these reasons, the standard error of the fit, which treats error bars as statistical uncertainties while they are here dominated by Type~B uncertainties, must not be given too large a significance. Eventually, one should bear in mind that the Type~B uncertainty of the $\nu_{\mathrm{Rb}}/\nu_{\mathrm{Cs}}$ determination is at best the Type~B uncertainty of the last 4 points: $4.4\times 10^{-16}$. To date, this is the second most accurate measurement (\emph{i.e.} measurement against PFSs) of an SRS, after the recent measurement of the 5s$^2$ $^1$S$_0$-- 5s5p $^3$P$_0$ transition in $^{87}$Sr \cite{LeTargat2013a}.


The fitted value of $\nu_{\mathrm{Rb}}/\nu_{\mathrm{Cs}}$ appears to have an offset with respect to the value based on the 2004 recommended frequency. This offset is well within the 2004 recommended uncertainty ($3\times 10^{-15}$ \cite{CCTF2004}\cite{CIPM2007}) shown by the grey area in Fig.~\ref{Fig_RbCsFinal}.

Data in Fig.~\ref{Fig_RbCsFinal} are also exploited to put new constraints on the variation of the fundamental constants in time and position, and on their couplings to gravity \cite{Guena2012a}\cite{Tobar2013}. They provided the first differential gravitational red shift test for the Rb and Cs systems.

\section{Contributing to TAI with the {\forb} frequency standard}\label{sec_TAI} %

\subsection{Including {\forb} calibrations in \textit{Circular T}} \label{sec_circularT}

\begin{table}
\begin{center}
\caption{Two examples of {\forb} calibration data reported to the BIPM, in January 2012 (\textit{Circular~T289}) and June 2013 (\textit{Circular~T306})
 \label{tab_Report}}
\begin{tabular}{ccccccc}
\hline \hline
Period&Mid-date&$y$&$u_{\mathrm{B}}$&$u_{\mathrm{A}}$&$u_{\mathrm{l/Lab}}$&$u_{\mathrm{SecRep}}$\\
\hline 
55924-55949&55936.5&-811.4&3.9&3.0&1.0&30\\
\hline 
56439-56469&56454&-1897.9&3.0&2.0&1.5&13\\
\hline \hline
\end{tabular}
\end{center}
{$y$ is the fractional frequency offset of $\mathrm{H}_{\mathrm{Ref}}$ determined with {\forb}, in units of $10^{-16}$. $u_{\mathrm{B}}$ is the Type~B uncertainty of the {\forb} secondary frequency standard, $u_{\mathrm{A}}$ the statistical uncertainty of the measurement, and $u_{\mathrm{l/Lab}}$ the link uncertainty (see Sect. \ref{subsec_DataTreatment}). The last uncertainty, $u_{\mathrm{SecRep}}$, is the uncertainty of the recommended value for the $^{87}$Rb hyperfine transition used as a secondary representation of the second. In the first example (Jan 2012), the 2004 recommended value was used: 6~834~682~610.904~324~Hz with a $3\times 10^{-15}$ recommended uncertainty \cite{CCTF2004}\cite{CIPM2007}. In the second example (June 2013), the revised value was used: 6~834~682~610.904~312~Hz with a $1.3\times 10^{-15}$ recommended uncertainty \cite{CCTF2012}\cite{CIPM2013}.}

\end{table}

Measurements presented in Fig. \ref{Fig_MaserFO2Rb} yield frequency calibrations of the LNE-SYRTE reference hydrogen maser $\mathrm{H}_{\mathrm{Ref}}$ (Fig. \ref{fig_metchain}) against the $^{87}$Rb secondary representation of the second realized by {\forb}. Such calibrations performed over the conventional 5-day grid open the possibility to be used to calibrate TAI. In order to pursue this possibility for a Secondary Representation of the Second to contribute to TAI,
we submitted a set of such calibrations to the BIPM in January 2012, and proposed that this submission be reviewed by the Working Group on Primary Frequency Standards (WGPFS) following the procedure usually applied for a PFS reporting for the first time.

This submission comprised a set of eight calibration reports based on measurements performed between January 2010 and December 2011, together with two publications \cite{Guena2010}\cite{Guena2012} on the {\forb} fountain and its uncertainties.
In addition to the three usual uncertainties assigned to calibrations with a PFS, a fourth uncertainty $u_\mathrm{{SecRep}}$ corresponding to the recommended uncertainty of the recommended value of the SRS is introduced. Our submission also included a note showing the January 2012 version of Fig. \ref{Fig_RbCsFinal} and pointing out that the recommended value for the $^{87}$Rb SRS had to be updated. In April 2012, the WGPFS recommended that the BIPM publish henceforth {\forb} calibrations in \textit{Circular~T}. Table \ref{tab_Report} gives the calibrations for the January 2012 and June 2013 periods as examples.
The WGPFS also recommended that the recommended value for the ground-state hyperfine transition of $^{87}$Rb be revised by the CCTF.

By November 2013, a total of 31 calibrations based on the $^{87}$Rb SRS realized with {\forb} were published in \emph{Circular~T}.

\subsection{Revised recommended value for the $^{87}$Rb ground state hyperfine frequency} \label{sub_sec_new_definition}

The 12$^{\mathrm{th}}$ point in Fig.~\ref{Fig_RbCsFinal}, which is based on measurements over the February 2012 to August 2012 period, was our most accurate determination of the $^{87}$Rb ground state hyperfine frequency at the time of the 19$^{\mathrm{th}}$ CCTF meeting (September 2012). This measurement gave $6~834~682~610.904~312$~Hz with an uncertainty $3~\mu$Hz, which corresponds to a fractional uncertainty of $4.4 \times 10^{-16}$. During this meeting, the BIPM also presented another determination based on {\forb} calibrations found in \textit{Circular~T} \cite{Petit2013}. This determination is largely independent given that the {\forb} data are covering a different period and that the primary reference is given by the whole TAI ensemble. The two determinations were in agreement, with a difference of less than 1 part in $10^{16}$. On that basis, the CCTF recommended a revised value for the $^{87}$Rb secondary representation of the second: $6~834~682~610.904~312$~Hz with a $1.3\times 10^{-15}$ recommended uncertainty \cite{CCTF2012}, corresponding to 3 times the uncertainty of our experimental determination, in accordance with the rules of the CCL-CCTF Frequency Standards Working Group. This revised value was adopted by the CIPM at its 102$^{\mathrm{nd}}$ meeting in June 2013 \cite{CIPM2013}. It is shifted by $-1.73 \times 10^{-15}$ with respect to the 2004 recommended value, a difference which is well within the 2004 recommended uncertainty of $3\times 10^{-15}$, as can be seen in Fig.~\ref{Fig_RbCsFinal}.

Note that since our submission of {\forb} data to the WGPFS in January 2012, this working group was renamed Working Group on Primary and Secondary Frequency Standards (WGPSFS).

\subsection{The {\forb} frequency standard \emph{versus} the TAI ensemble}

\begin{figure}
\includegraphics[angle=0,width=\linewidth]{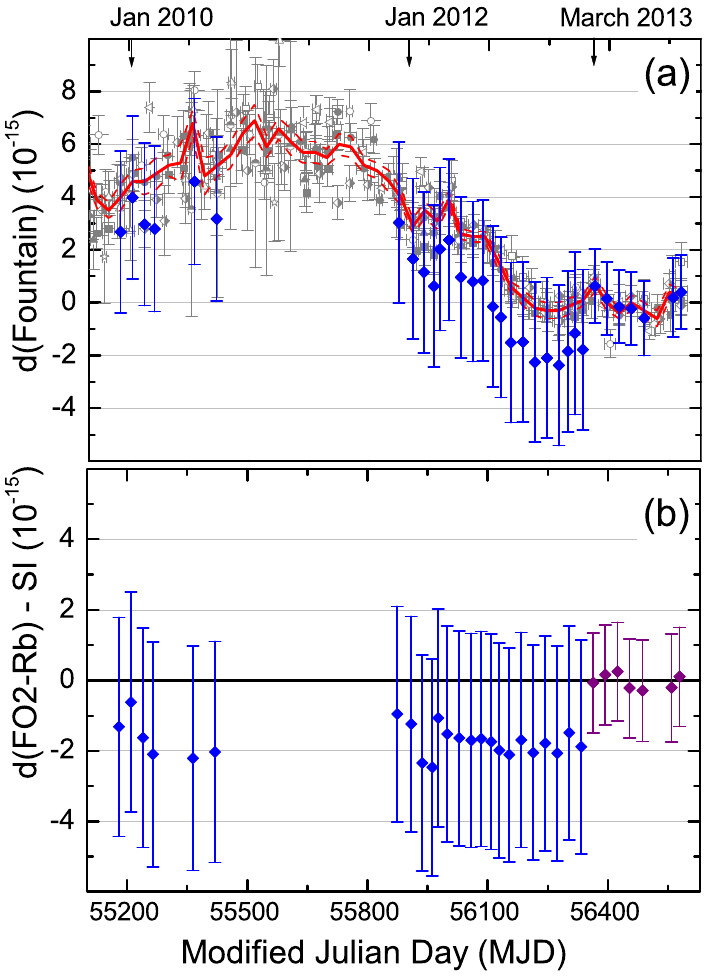}
\caption{Data extracted from \emph{Circular~T}. (a) The red curve (and red dashed lines) show the deviation $d$ (and uncertainties) of the TAI scale interval, as defined in section 4 of \emph{Circular~T}. It is estimated by the BIPM on a monthly basis based on individual PFS calibrations shown in grey. Blue diamonds are the estimation of $d$ by the BIPM based on the {\forb} calibration reports. The error bars include the recommended uncertainty for the $^{87}$Rb hyperfine transition used as a secondary representation of the SI second, $u_{\mathrm{SecRep}}$. Up until February 2013 included, {\forb} calibrations are made using the 2004 recommended value \cite{CCTF2004}\cite{CIPM2007}. For the 6 last points, {\forb} calibrations are made using the 2012 revised recommended value \cite{CCTF2012}\cite{CIPM2013}. (b) From (a), differences between d({\forb}) (from {\forb} calibrations) and $d$ based on the TAI ensemble, and corresponding uncertainties. The error bars also include $u_{\mathrm{SecRep}}$.}
\label{Fig_TAI}
\end{figure}

Figure \ref{Fig_TAI} presents comparisons of the {\forb} SFS to the TAI ensemble based on data extracted from \emph{Circular~T}. 
The red curve in Fig.~\ref{Fig_TAI}(a) shows the duration of the TAI scale interval $d$ as defined in section 4 of \emph{Circular~T}, estimated by the BIPM based on individual PFS calibrations, which are shown in grey. The red dashed lines display the standard ($1~\sigma$) uncertainty of this estimation determined by the TAI algorithm. Blue diamonds show the estimation of $d$ by the BIPM based on our {\forb} calibration reports. The error bars display the overall $1~\sigma$ uncertainty directly deduced from \emph{Circular~T}
by summing in quadrature the standard uncertainty $u$ estimated by the TAI algorithm for {\forb} and the uncertainty $u_{\mathrm{SecRep}}$ of the $^{87}$Rb SRS. Figure \ref{Fig_TAI}(b) shows the difference between {\forb} calibrations and $d$ from the TAI ensemble, and the corresponding uncertainties. {\forb} calibrations are all in agreement with the TAI ensemble within the standard uncertainties. In both graphs of Fig.~\ref{Fig_TAI}, the change of recommended value, applicable since March 2013, is clearly visible, as is the reduction of the recommended uncertainty. This change does not have a physical origin but is instead associated with the standardization process.

\begin{figure}
\includegraphics[angle=0,width=\linewidth]{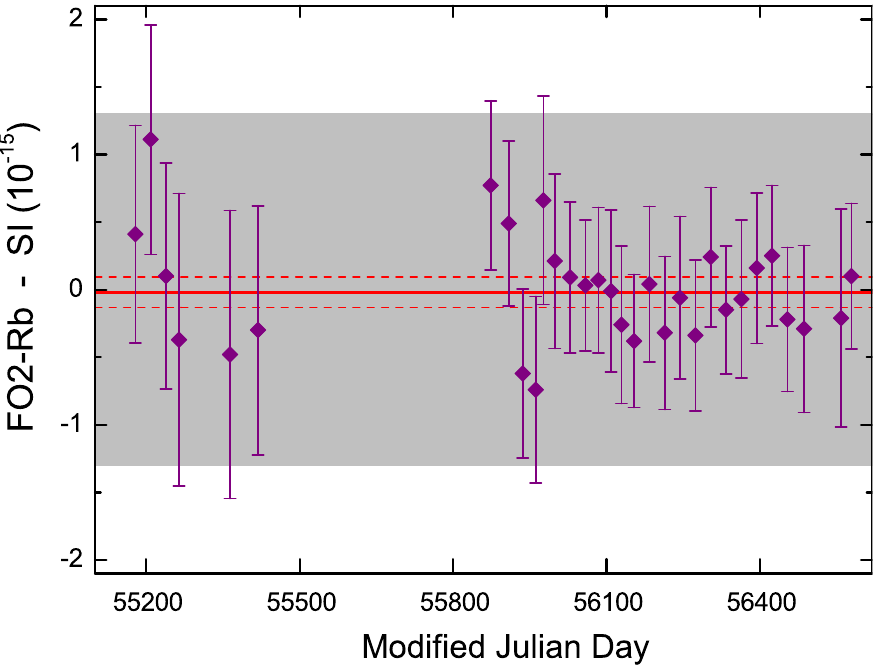}
\caption{{\forb} frequency standard \emph{versus} the TAI ensemble. This plot is obtained by taking the data points of Fig.~\ref{Fig_TAI}(b) and shifting the blue points in such a way that, in effect, all points are referenced to the same value for the $^{87}$Rb hyperfine transition, namely the 2012 recommended value \cite{CCTF2012}\cite{CIPM2013}. Here, error bars do not include the recommended uncertainty $u_{\mathrm{SecRep}}$. Instead, this uncertainty is represented by the grey area. The red lines show the result of the fit of the data points to a constant used to test the consistency of the data (see text).}
\label{Fig_DiffRbSI}
\end{figure}

A physical view of actual frequency variations between the {\forb} SFS and the TAI ensemble is shown in Fig. \ref{Fig_DiffRbSI}, where data are displayed now using the same reference value for all points, and with error bars showing $u$ only. The consistency of this data is tested with a weighted linear fit to a constant. We obtain $\chi^2=0.34$, $Q=1.0$ and $R_{\mathrm{B}}=0.58$ which indicates highly consistent data set. The fitted constant is $-0.01\times 10^{-16}$ and the standard error of the fit is $1.1\times 10^{-16}$. This result, in effect, constitutes an absolute frequency measurement of the {\forb} SFS directly against the TAI ensemble. This analysis demonstrates that it is possible to perform such an absolute frequency measurement with a statistical uncertainty at the level of 1 part in $10^{16}$.

The above analysis is solely based on data publicly available in \emph{Circular~T}. Similarly to what was done for comparing PFSs 
\cite{Parker2012}, \cite{Petit2013}-\cite{Parker2010},
 a refined treatment can be obtained via the calculation of TT(BIPM) \cite{Petit2013}.

\emph{Note~1}-- An independent $^{87}$Rb \emph{vs} $^{133}$Cs fountain comparison was reported for the first time in 2010. Agreement with LNE-SYRTE measurements is claimed at the $10^{-15}$ level but, to our knowledge, the actual result of the measurement has not yet been published \cite{Ovchinnikov2011}-\cite{Ovchinnikov2011a}.

\emph{Note~2}-- Several Rb fountains are operated at the US Naval Observatory (USNO) 
\cite{Peil2006}-\cite{Peil2013}. No uncertainty evaluation is published for these fountains. Nevertheless, comparisons of these fountains with the {\forb} SFS could be valuable for testing the stability of remote time and frequency transfer methods and of {\forb} and the USNO fountains themselves.

\section{Conclusions}

In this paper, we reported on a Secondary Frequency Standard which is capable of performing as well as Primary Frequency Standards in all their typical applications. With this SFS, we have, for the first time, experimented with all the necessary steps for an SFS to participate in international timekeeping, within the framework of the CIPM and its committees and working groups. As part of this experimentation, we have observed the effect of the standardization process consisting in defining and revising the recommended values for Secondary Representations of the Second, when such an SRS is actually used for calibrations. We reported on a highly accurate absolute frequency measurement of the SRS used in this work, \emph{i.e.} the $^{87}$Rb ground state hyperfine splitting, with a total uncertainty of $4.4\times 10^{-16}$. This measurement was obtained by comparing the {\forb} atomic fountain frequency standard to LNE-SYRTE PFSs. We have also shown that such an absolute frequency measurement can be done directly against the TAI ensemble, with a statistical uncertainty of 1 part in $10^{16}$, by direct use of data published in \emph{Circular~T}. This exemplifies the dissemination of the SI second via TAI at the accuracy limit of PFSs.

As a final step, calibrations provided by the {\forb} SFS are now used to determine steering corrections of TAI, starting with \emph{Circular~T307}. This is the first time that a Secondary Frequency Standard contributes to steering TAI.

In the future, the rapid progress of frequency standards based on optical transitions will place several of these standards in the position to also contribute to TAI. We believe that the present work has contributed defining and clarifying several aspects of this process, and triggered useful work to anticipate the likely situation of diverse SFSs contributing together. We hope that this will be a useful contribution toward a possible redefinition of the SI second based on optical transitions.

\section*{Acknowledgments}

We are grateful to the members of the BIPM, of the WGPFS/WGPSFS, of the CCL-CCTF Frequency Standards working group and of the CCTF, and their chairmen, for their work and for the kind attention given to our material.

We are grateful to the University of Western Australia and to M.E.~Tobar for the long-lasting collaboration which gives us access to the CSO. We are grateful to the SYRTE's technical services for their continued contribution to this work. We thank J. Lodewyck, L. Lorini, P. Tuckey and P. Wolf for their comments and contributions to the manuscript.

This work is conducted at SYRTE: SYst\`{e}mes de  R\'{e}f\'{e}rence Temps-Espace. It is supported by Laboratoire  National de M\'{e}trologie et d'Essais (LNE), Centre  National de la  Recherche Scientifique (CNRS), Universit\'{e} Pierre
et Marie Curie (UPMC), and Observatoire de Paris. LNE is the French National Metrology Institute.



\end{document}